# Faceted Semantic Search for Personalized Social Search

**Massimiliano Dal Mas**

*me @ maxdalmas.com*

**ABSTRACT**

Actual social networks (like Facebook, Twitter, Linkedin, ... ) need to deal with vagueness on ontological indeterminacy. In this paper is analyzed the prototyping of a faceted semantic search for personalized social search using the "joint meaning" in a community environment. User researches in a "collaborative" environment defined by folksonomies can be supported by the most common features on the faceted semantic search.
A solution for the context-aware personalized search is based on "joint meaning" understood as a joint construal of the creators of the contents and the user of the contents using the faced taxonomy with the Semantic Web.
A proof-of concept prototype shows how the proposed methodological approach can also be applied to existing presentation components, built with different languages and/or component technologies.

## INTRODUCTION

Web allows to accumulate digital data at an unprecedented rate. Information overload is an ongoing problem that diminishes quality of data among knowledge. Folksonomies are a set of terms that a group of users tagged content without a controlled vocabulary, that lead to a number of limitations and weaknesses in folksonomies.

This paper wants to explorer the use of the pragmatics of dialogue to constitute dynamic connections between concepts of knowledge in a collaborative environment. To develop this proposal it was necessary to improve the quality of the individual tags on folksonomies given a classification order according to different facets. Folksonomies were integrated with faceted navigation to balance the rigidity of facets built from a controlled vocabulary with the potential anarchy of raw folksonomies.

--------

This work considers a mechanism based on the "joint meaning" [1], as commitment providing a way to connect speech acts from who contribute to the web, called as *speakers*, and who search the web, called as *reader*. Joint meaning will be used as a procedure to deal with vagueness on ontological indeterminacy [2] given the ability to extract semantic faceted metadata and create semantic associations leading to better search [3], integrate and analysis on folksonomies [4].

The system will allow dynamic selection of categories using better an auto-classification developed integrating the Rich Internet Application features (RIA) of the Web 2.0 with the meaning of the Semantic Web - Web 3.0 [5].

The contributions of this paper are as follows:
- it introduces the use of the concept of joint meaning for faceted folksonomies
- it shows the approach proposed for a faceted dynamic design
- it describes the architecture and the implementation of the developed prototype
- finally, it draws conclusion and suggest directions for future works

**BACKGROUND**

**Faceted Classification**

In social networks, considered as a subset as Web 2.0 if not a related concept, new knowledge is created by sharing ideas as information done by data. Semantic Web for the social network can help in sharing information and knowledge creating communities of people with similar interests.

A predefined structure to organize information is not applicable for a social network that can be represented as a graph in evolution of a collaborative environment where users can create different kinds of classification on the fly from the previously decided. It is so necessary to assign an item to multiple parameters each representing an aspect or a "facet" of the information.

In such scenario a new springtime arose on the concepts of "facets" and "faceted classification" as multidimensional classification developed by Ranganathan in the context of classical librarianship.

The faceted classification of an object, created by exploiting a system of attributes (metadata) representing each one aspect or property , is be able to describe exhaustively the object itself.

Considering a faceted classification only as a theoretical apparatus coined by science books is limitative. This approach, in fact, is the formalization of a technique of communication that we often use in a wide range of contexts, from the organization of personal information. [6, 7, 8, 9]

The faceted classification has important advantages over other systems in particular: *multidimensionality*, *persistency*, *flexibility* and *scalability*. These features prevent the deterioration of repository avoiding that changes have negative repercussions on the information organization.

Facets can support web search with a dynamic personalization's, multiple interests and multiple info behaviors. An integrated search-browse-facet user interface provides simple complexity supporting both quick answers to specific questions and deep research exploration.

Facets are orthogonal with mutually exclusive dimensions (for example: a seminar is not a person is not a document and is not a place). Facets could be an important addition to search / browse if used with an active interface in a dynamic combination of search and browse applied at search time with a post-coordination and not a pre-coordination, as in the advanced search. But facetted classification requires adding lots of metadata and to understand users for the information architecture.

Faceted interfaces result more intuitive for its simplicity of internal organization that allows multiple perspectives with the ability to handle compound subjects.

An internal facet structure reflects current usage. Users of a community can understand different kind of information structures matching the same domain and task.

Having a precision of unit values facets allow flexibility for additions of new subject, facets, or entities at any point in the system with different kind of categorization (e.g.: chronological, alphabetical, spatial, simple to complex, size or quantity, hierarchical, canonical). The implementation of faceted classification as user interface usually have some disadvantages: difficult to define the browsing scope, loss of context, difficult to state complex relationships, limited type and size on domain applicability.

**Folksonomies**

Folksonomies, used in Web 2.0, are building bottom-up classification systems but they are not a classification system. They are an unordered, flat set of keywords that are ranked by popularity. Tag clouds are actually not organized by a ranking criterion at all and only highlight terms based on their frequency. That can tell you how groups of people are thinking, but it does not tell you anything about the relationships between words or concepts because there is not a system of rules [10, 11, 12].

Between folksonomies there are conceptual relationships expressed when two or more web sources are tagged with the same tag and these relationships can grow in very complex ways revealing a great deal about how people think and how some ideas can be related. Browsing through tags it is possible to see how other people have tagged a particular set of web sources (as articles, comments, pictures, video, etc.) and thus their conceptual relationships.

But it's not clear how the overall set of connections between concepts constitute an organization of knowledge. As the number of tags and sites is going to increase the complexity of the tags and community relationships grows exponentially and the lack of a system of rules becomes particularly clear.

Folksonomies do not organize information, but aggregate individual acts of cataloging ranked by popularity. So folksonomies cannot be compared with taxonomies, thesauri, or ontologies because they are not a classification system.

**FACETED SEARCH AND PERSONALIZED SEARCH**

**Joint Meaning**

Every kind of system based on a search engine tries to give the right answer to the user, but what is the meaning of the user query? And what the user have in mind?

In this work the act of search was considered as a communicative act between the users of a community performed by a search engine with certain types of actions guided by the joint meaning here described.

According to the Speech Act Theory [13, 14] what a speaker wants to communicate depends on his/her intention being a function of that. The meaning of a linguistic expression in a language is determined by how it is used by a community of people. For a holistic view, the meaning involves connection between inferential and evidential connections and actions that people could take under various circumstances.

As stated by Herbet Clark the meaning is jointly constructed by the *speakers* and the *reader*. Community environments, like Facebook, are bringing the web on a more *Communicative Act* between a *speakers* and his/her *reader*.

The meaning of the communicative act produced by the users of a community appears to be collectively constructed by the *speakers* (as users that contribute to their own community space) and by the *reader*, called by some communities as "friend" (that can see the speakers community spaces).

We can consider $U = \{1, ..., n\}$ as a finite set of *n users*, supposing that $n–1$ users of $U$ communicate something to the *user u*. We can designate $u$ as the *reader*, and the other $n–1$ users as the *speakers U–{u}*.

While speakers meaning was solely a function of the speakers communicative intentions, joint meaning was a collective construal of the *reader u* and the *speakers U–{u}*. This work considers meaning as a conflict between reader meaning, understood as a personal mental state, and a collective construction between the personal mental as joint meaning. Every kind of meaning was understood as a different facet of knowledge, so making something common between the speakers and the reader means having a common faceted visualization and it means communicate. For the proposed work joint meaning can be considered as joint activities of two or more subjects' users that can develop together a *faceted interface $f_n$* according to their communication.

Communication can be defined by the fixpoint axiom of mutual belief that we can considered as the *fixpoint axiom of faceted interface for the communication*. In some cases the joint meaning of a communicative act from the *reader u* can coincide with one *speakers meaning* of the *speakers U–{u}*. But in many cases joint meaning can be different from the reader meaning, whether or not the communicative intention of *speakers U–{u}* has been correctly understood by the *reader u* we have to deal with vagueness on ontological indeterminacy used by the *reader u*.

The Semantic Web developers encodes all the information in an ontology filled with rules that say, essentially, that "Robert" and "Bob" are the same. But humans are constantly revising and extending their vocabularies, for instance at one times a tool might know that "Bob" is a nickname for "Robert," but it might not know that some people named "Robert" use "Rob", unless it is told explicitly.

Conversations could be seen as sequences of communicative acts between one or more *speakers*, each of which has an associated speaker's meaning that depends on his/her communicative intentions, and a *reader* with his/her "understanding" represented by an own ontology.

*Joint meaning* is formed every time by a *reader u* and the *speakers U–{u}* performing functions to maintain a "shared view" of what it is said. It is not just a common belief of what has been said, caused it may not coincide with the original speaker's meaning, but it is more a joint commitment of two or more subjects, who are obligated to each other to act coherently caring out deontic implications.

Suppose for example that Mark, our *reader u*, asks to a semantic search engine, see Figure 1:
"*I think I'm searching for a sport car*"
with the intention to search a faster and style car having in mind a Ferrari and imagine that the semantic search answers:
"*The best sport utility vehicle (SUV) is a Hummer H1*"
Clearly the search has taken up Mark's statement as a search of an off-road vehicles according to the actual trend connection between sport car and SUV using the relative semantic relation between the asking for a sport car, tag $t_n$, and an aspect of the knowledge of SUV, facet $f_n$, expressed by *($t_n, f_n$)* as a sport car *R* depicted in (1).

$$(1) \quad R = (t_n, f_n)$$

This semantic relation was determined by a speaker, of the *speakers U–{u} group*, that considers a SUV as a sport car and it is expressed by the relation *($t_n, f_n$)*.

Mark may now search answer, thus implying that his original statement was on searching a trendy car, by asking for instance:
"*Pity. Well, show me different body styles of SUV*"
So he is refined his search on body styles faces of the SUV knowledge

At this point of the search the joint meaning of Mark's original statement is turned out as a search for a SUV car. Then suppose that the semantic search engine answers:
"*A Maserati Kuba SUV*"

The initial meaning of the Mark utterance was to search a car like a Ferrari, but now without violating a joint commitment he has accepted, as a matter of joint meaning, that such an utterance was a search on a trendy car, and he is acting coherently to the search engine answers.
So he is taking the facet of the trendy car folksonomies from the sport car knowledge according to the joint meaning using the semantic relation between sport car and trendy car. This relation can change during the time, for instance in a close future a trendy car could become a van.

Evolution of language that confounds parents who can't understand the slang of their teenagers also can trips up these systems. Folksonomies tags could be used to follow the evolution of the language to conceptualise different facets of a common knowledge to express users' preferences and needs since they allow users to add their own tags based on their interests.
Overwhelm criticism that folksonomy tags are ambiguous and uncontrolled terminology using more facets for a tag it will possible to reflect real users' views and their vocabulary.

Joint meaning procedure can help to express what the user have in mind to have a personalized answer from a semantic search engine.

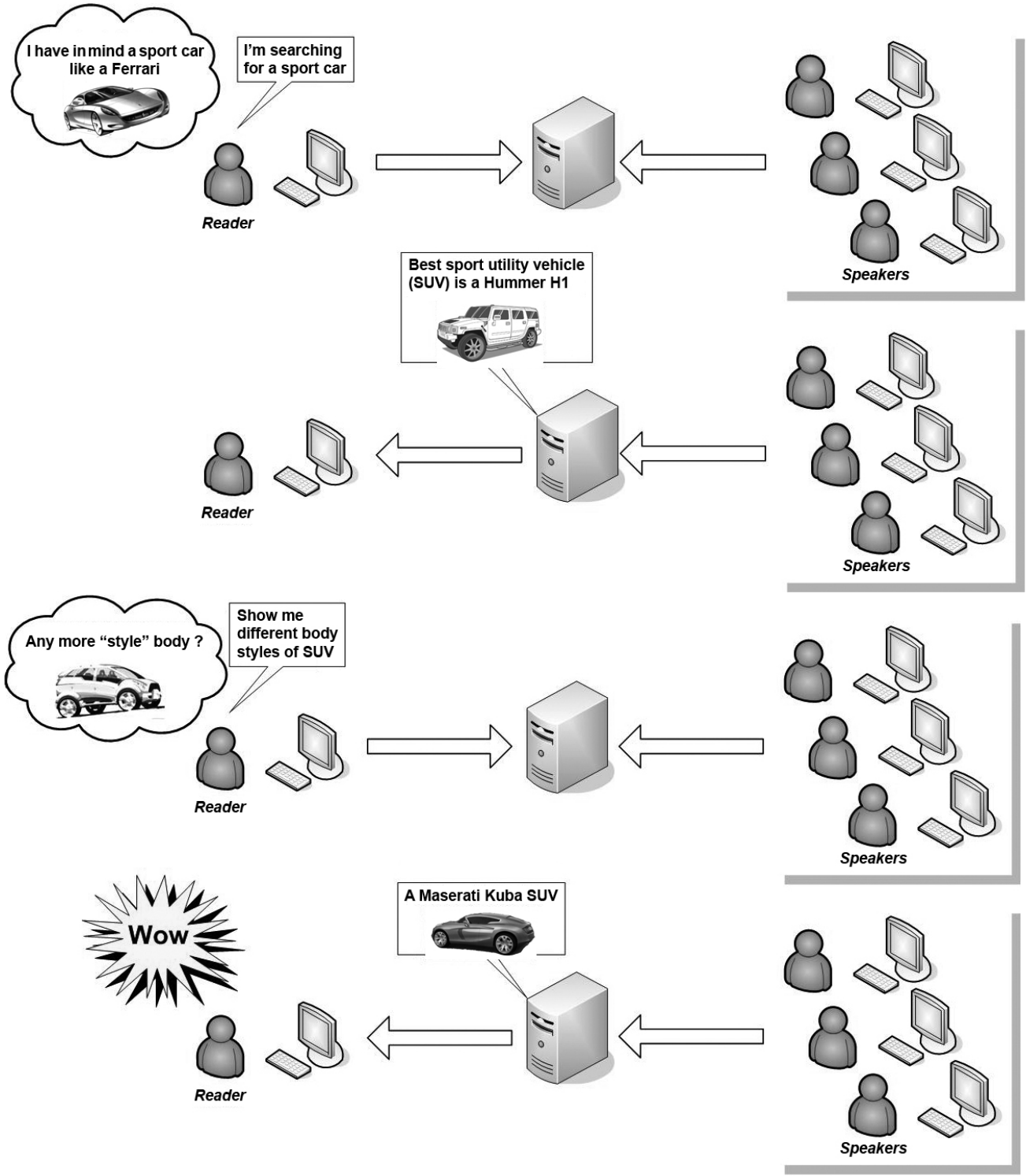

*Figure 1. Conversations could be seen as sequences of communicative acts between one or more speakers, each of which has an associated speaker's meaning that depends on his/her communicative intentions, and a* reader. *A semantic search engine makes the* joint meaning *J, expressed by (3) between the* reader *u and the* speakers *U–{u}*

For the joint meaning procedure we can use a set of pairs *(tn, fn)* to represent a faceted taxonomy as stated in (1) extended to all the *tn tag* of the folksonomies and the *fn* relative *facet*.

$$(2) \quad R = \{(t_n, f_n), \dots (t_n, f_n)\}$$

The multiple association *R* consists of a number of concepts that should be equivalent with each other expressed by facets *fn* and the relative tag *tn* matching from the folksonomies defined by the *speakers U–{u}*.

The joint meaning *J*, expressed by (3) between the *reader u* and the *speakers U–{u}* towards a semantic search engine, will be defined as a multiple association defined using a *Semantic Enrichment Method*. It consists of a number of a multiple association of concepts *R*, defined by the *speakers U–{u}*, and a domain ontology *O* used by the *reader u* that depend on the context of a user. As ontology *O* was used the translation of the Suggested Upper Merged Ontology (SUMO) into OWL [15] adding concepts for domain specific content that wasn't well supported at the upper level.

$$(3) \quad J \subseteq R \times O$$

To disambiguates multiple matching was used the *Superconcept Formation System (SFS)* [16], a learning-based matching algorithm combined with rule based techniques that uses a n-dimension Euclidean space vector for concepts with one semantic aspect; for each dimension was used an Artificial Neural Network technique to learn the relative weight and was calculated the weighted sum of dissimilarities from all corresponding dimensions.

**Implementation**

Created for people community the prototype website has been developed locally on the common functionalities of actual social networks (as Facebook, Myspace, Linkedin, Twitter, …) for testing the proposed approach. For the implementation was used the popular CMS Drupal with a plug-in that enables RDF and OWL output [17], and a themed AJAX interface used to retrieve data integrating Flex.

Faceted interface was defined in the Semantic Web by triple to define the elements properties composing. Those were stored into the database access system, a SPARQL engines was integrated using RIA to combine the collaborative nature of Web2.0 with the ontologies of the Semantic Web (Web 3.0).

Every faceted interface was composed by different kind of objects identified by means of more than one aspect. So every object had more than one facet and the composition of every chosen facet composed the faceted interface between the *reader* and the *speakers* that could be identified by means of another facet. The faceted interface had a *joint meaning* in multiple domains as a reference designation of the interface with respect to the *speakers* and the *reader* being related to facets, see Figure 2.

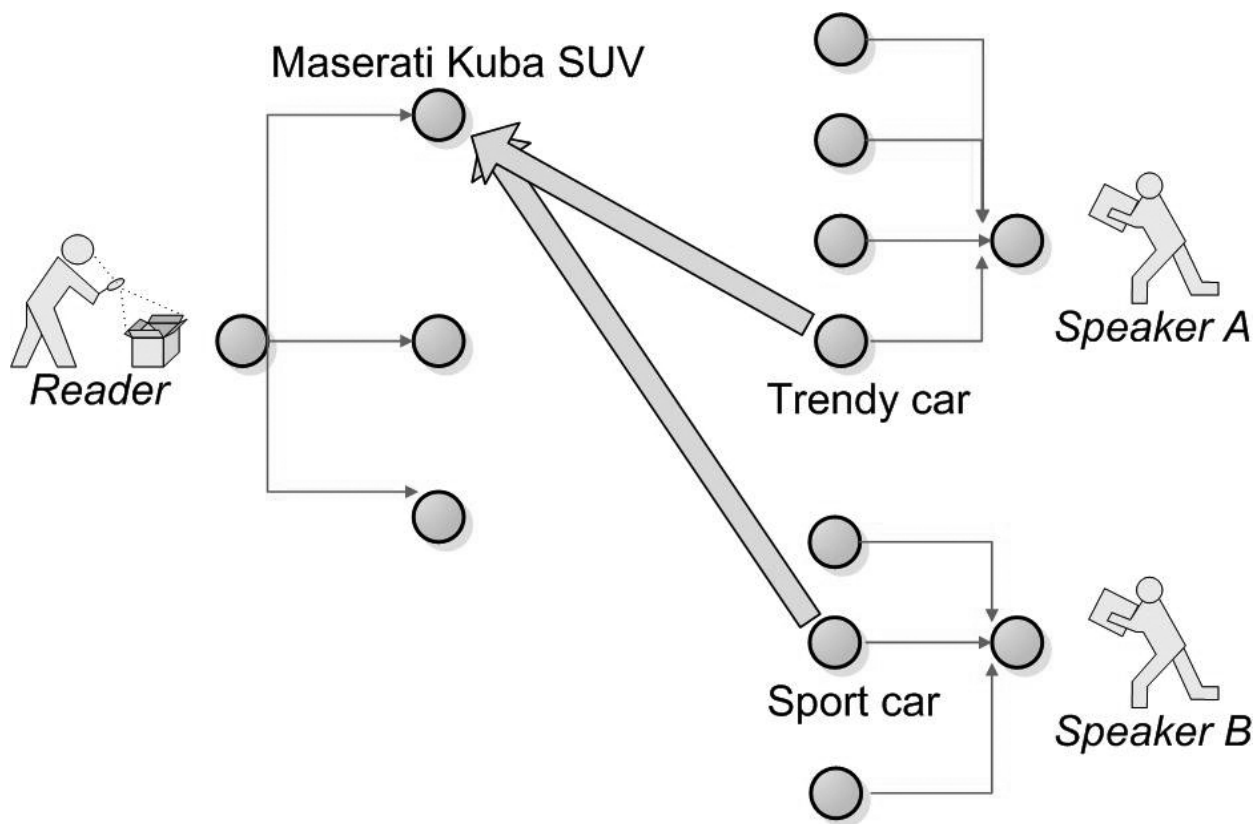

*Figure 2. Faceted multiple matching: the system may aggregate some faceted elements into "Maserati Kuba SUV" tag for the* reader *taking the "Trendy car" facet from the* speaker A *and the "Sport car" facet from the* speaker B.

A centralized management of the identification register was used for the objects. Using the Semantic Web the information referred to any object can be arbitrarily voluminous and structured, having any desired information granularity. Being flexible it is not required the use of long identification. So the identification can easily be kept stable over time; while at the same time the content of the metadata can be adapted to current needs (i.e.: restructured, increase of granularity). The information could be fragmented or assembled as needed including graphical presentations [18].

Instead of "smart" economizing with computing power it had become more essential to describe things logically and straightforward in order to enhance functionality, exchangeability and communicability. Moreover the reference designations should be used over the entire life cycle of the "objects".

The faceted interface was constructed using algorithm based on the joint meaning. Ranganathan's theory could help us to automatically determine intuitive facets that belong to either of intuitive and unintuitive categories.

Acquisition of faceted subject metadata was done by a *folksonomy social tagging* used as a means towards building such structure consolidated towards the Semantic Web. Folksonomy were related to every kind of contribution of the users (*speakers*) like text, picture, audio, video or free code that were considered as a *portlet* framework. These portlets were made either manually, writing code by the user, or via a portlet creation wizard.
A SUMO ontology, with adding concepts for domain specific content, was developed to contain the knowledge that was supposed to be used by the user of the contents (*reader*).

**Related work**

In literature there are a lot of works on identifying assertions from the web [19] extracting information from the web database [20, 21] and researches in the area of "semantic search" to improve search and navigation by combining ontologies and social tags [22] nevertheless this paper aims to propose some preliminary ideas on how to create taxonomies of facets using the "joint meaning" for attributes and facets.
The use of "joint meaning" to design personalized search in collaborative environment is an underexplored area that arose from the pragmatics [1].
Different kinds of tools have been developed for different aspects of the work. Two main classes are summarized below:

- Faceted Navigation

Most faceted navigation are used as interface for searching a large content database [23] not considering different kinds of faceted visualization according to different kinds of *readers* (as done with "joint meaning") and using a fixed matching algorithm instead of a learning-based matching algorithm for automatic ranking of facet quality [24] (as could be seen in the work of Diederich, Balke, and Thaden)

- Social Networks

Nowadays the Semantic Web is looked by social networks for providing their bones. Adobe and Facebook cooperate to allow web developers to create RIA applications [25] with the opensource framework Flex [26]. Facebook is working to use the Semantic Web on social network data [27] used to predict some individual private trait, as could be seen in the work [28]. Outside of Facebook, *"microformats"* exist for tagging all kinds of information on ordinary web page [29]. Microformats are simple standards that add significance in HTML documents just let you mark up HTML elements, to make accessible information to third party software and services.
The official microformats site contains the development information and scripts to create the "*major microformats*".
Social Networks as Facebook provides features for listing the people you know, publishing contact information, and advertising planning events for group of friends not allowing a personalized search from the *speakers* contributions and the *single reader* as done using the joint meaning.

**Evaluation**

The evaluation was based on MiLE+ (Milano Lugano Evaluation Method) [30], which proposes an approach to usability evaluation under application-independent analysis (based on usability

TABLE I
EXAMPLE OF EVALUATION MATRIX

| Task: Find a sporting elegance car with a sophisticated technology | PREDICTABILITY | UNDERSTANDABILITY | RICHNESS | COMPREHENSIBILITY | GLOBAL SCORE FOR THIS TASK |
|---|---|---|---|---|---|
| Scores | 8 | 8 | 5 | 6 | 6.75 (average score) |
| Weights | 0.1 | 0.1 | 0.5 | 0.3 | |
| Weighted Scores | 0.8 | 0.8 | 2.5 | 1.8 | 5.9 (weighted score) |

*Only readers with same interest in a sport trendy cars can find the Maserati Kuba SUV*

principles done by different experts) and application-dependent analysis (based on the requirements of the application, providing a step-by-step action guide for detecting the different problems with an assigned task). See an example on Table I where it is depicted the predictability, understandability, richness and comprehensibility of the personalized faceted search interface designed according to the "joint meaning".

The overall results confirm that people (on a sample of 150) prefer the faceted interface (83%), finding it useful (95%) and easy-to-use (82%).

**CONCLUSION**

In this paper has been shown the use of the linguistic pragmatics in faceted search to deal with vagueness on ontological indeterminacy expressed by the user search (the *reader ontology O*) for personalized social search in a "collaborative" environment defined by folksonomies (multiple association of concepts $R$ expressed by facets $f_n$ and the relative tag $t_n$ ).

The main goal of this article has been to explore the basic argument on faceted search for folksonomies and useful aspects of pragmatics of dialogue towards the "joint meaning" understood as a joint construal of the creator of the community contents (*speakers*) and the user of the community contents (*reader*) thanks to the context adaptation using a faced taxonomy with the Semantic Web.

A prototype based on the proposed methodology was implemented to test its implementation by the actual technologies and its evaluation by a sample of users.

The ability to search on tags contributed by users with no a priori structured knowledge will be a crucial requirement of the information architectures that will make it possible to search in a personalized context-aware.

The described work has only scratched the surface of a huge problem being an initial step of a research program that will address several open issues:
- a deeper comprehension of social commitment by subjects interaction as social reality intentionally constructed and how deontic affordances could be considered to produce joint meaning (see Carassa & Colombetti, 2009, for a first step in this direction [1]);
- enriching the methodological approaches by considering in depth different kind of RIA behaviours like: chat, multimedia synchronization, etc. [31,32];
- developing automatic metrics for automatic facet ranking from the *Superconcept Formation System (SFS);*
- working on the semantics of the conceptual model, to enable automated methodological approaches;
- continuing the industrial experimentation, by targeting different kinds of platforms.

## ACKNOWLEDGMENT

I would like to thank Professor Marco Colombetti for his advice on Knowledge Engineering. I am especially indebted to all the reviewers' detailed comments and constructive suggestions on the manuscript.

**Massimiliano Dal Mas** is an engineer at the Web Services division of the Telecom Italia Group, Italy. His interests include: user interfaces and visualization for information retrieval, automated Web interface evaluation and text analysis, empirical computational linguistics, text data mining, knowledge engineering and artificial intelligence. He received BA, MS degrees in Computer Science Engineering from the Politecnico di Milano, Italy. He won the thirteenth edition 2008 of the CEI Award for the best degree thesis with a dissertation on "Semantic technologies for industrial purposes" (Supervisor Prof. M. Colombetti).